\documentclass[twocolumn,floats]{revtex4}
\usepackage{graphicx}
\usepackage{amssymb}
\newcommand{\upcite}[1]{\textsuperscript{\cite{#1}}}

\begin{document}
  \title{Transient transition from free carrier metallic state to exciton insulating state in GaAs by ultrafast photoexcitation}
  \author{X. C. Nie$^{1}$}
  \author{H. Y. Liu$^{1}$}
  \email[Corresponding author: ]{haiyun.liu@bjut.edu.cn}

  \author{Xiu Zhang$^{1}$, Peng Gu$^{1}$, Hai-Ying Song$^{1}$, Fan Li$^{2}$,
  Jian-Qiao Meng$^{3}$, Yu-Xia Duan$^{4}$}

 \author{Shi-Bing Liu$^{1}$}
  \email[Corresponding author: ]{sbliu@bjut.edu.cn}

\affiliation{
\\$^{1}$Strong-field and Ultrafast Photonics Lab, Institute of Laser Engineering,
Beijing University of Technology, Beijing 100124, P. R. China
\\$^{2}$Department of Chemistry and Chemical Engineering, College of Environmental and Energy Engineering,
Beijing University of Technology, Beijing 100124, P. R. China
\\$^{3}$Hunan Key Laboratory of Super-microstructure and Ultrafast Process, School of Physics and Electronics,
Central South University, Changsha, Hunan 410083, P. R. China
\\$^{4}$School of Physics and Electronics, Central South University, Changsha, Hunan 410083, P. R. China
}

\begin{abstract}
We present systematic studies of the transient dynamics of GaAs by ultrafast time-resolved optical reflectivity. In photoexcited
non-equilibrium states, we found a sign reverse in transient reflectivity spectra $\Delta R/R$ (t $>$ 0), from positive around
room temperature to negative at cryogenic temperatures. The former corresponds to a transient free carrier metallic state, while the
latter is attributed to an exciton insulating state, in which
the transient electronic properties is mostly dominated by excitons, resulting in a transient metal-insulator transition (MIT).
Two transition temperatures (T$_1$ and T$_2$) are well identified by analysing the intensity change of the time-resolved
optical spectra. We found that photoexcited MIT starts emerging at T$_1$ as high as $\sim$ 230 K, in terms of a negative dip feature
at 0.4 ps, and becomes stabilized below T$_2$ associated with a negative constant after 40 ps in spectra. Our results address
a phase diagram that provides a framework for MIT through temperature and photoexcitation, and shed light on the understanding
of light-semiconductor interaction and exciton physics.
\end{abstract}
\date{\today}
\maketitle
\section{Introduction}
Metal-insulator transition (MIT) has been of great interest for the past half century in different branches of condensed matter
physics, ranging from semiconductors to strongly correlated electronic systems\upcite{{Mott-IMT,Mott-IMTs}}. It is of great value
in fundamental quantum many-body physics in solids\upcite{Chemla-Nature-2001}. By implementing ultrafast photoexcitation,
a growing interest has been devoted to light-induced MIT in strongly correlated systems, through melting of
charge-density-wave (CDW) in low-dimensional CDW crystals\upcite{ATCDW-PRL-2009,JC-PRL-2011,HY-PRB-2013},
resonant phonon-driven in manganites\upcite{DP-NM-2007,RIR-PRL-2008}, etc. In semiconductors, transient MIT can be induced
by the interplay between free carriers and excitons, for example, a exciton-dominant abrupt change was experimentally found in
photoluminescence dynamics below 49 K in GaAs\upcite{Amo-PRB-2006}. Further researches have explored the formation
dynamics of excitons and exciton Mott transition in GaAs quantum wells (QW)\upcite{{QiZhang-2016,Kaindl-2003}},
Si\upcite{{Suzuki-PRL-2012,Suzuki-PRL-2009,Suzuki-PRB-2011}} and Ge\upcite{Sekiguchi-PRB-2015} by using the technique
of optical-pump THz-probe spectroscopy.

GaAs is a typical direct band-gap semiconductor with a face-centred cubic (FCC) Bravais
lattice\upcite{{Blakemore-JAP-1982},{Harrison-FCC-2016}}, as shown in Fig. 1(a). Both the top of the valence band (TVB) and the
bottom of the conduction band (BCB) lie at $\Gamma$. It is well known that exciting GaAs by near-infrared (NIR) photons above the
band-gap around room temperature excite hot free carriers including free electrons in TVB and holes in BCB, resulting in a
transient metallic state. Unlike free carriers, excitons are pairs of electrons and holes in semiconductors, which can be operated
in optoelectronic devices, for example, excitonic integrated circuit for signal processing and communication through the direct
transform between photons and excitons\upcite{AAH-Science-2008}. It is of critical importance to conduct comprehensive
investigation on the interplay between photons, free carriers and excitons, and their contributions to physical properties.

In this paper we present systematic studies of the dynamics of photoexcited free carriers and excitons in bulk n-type GaAs(100),
in a wide range of lattice temperatures and excitation densities. We observe a transient MIT in the system, i.e., from free carrier
metallic state to exciton insulating state, by decreasing temperature. The exciton insulating state starts emerging
and gets stabilized at unexpected temperatures as high as $\sim$ 230 K and $\sim$ 180 K, respectively. In addition, our results
reveal that photoexcitation can effectively induce MIT in semiconductors, paving the way for the control of MIT and the understanding
of exciton physics.

 \begin{figure}                                                                                                                                             
  
   \includegraphics[width=1\columnwidth,scale=1,bb=0 0 2648 2284]{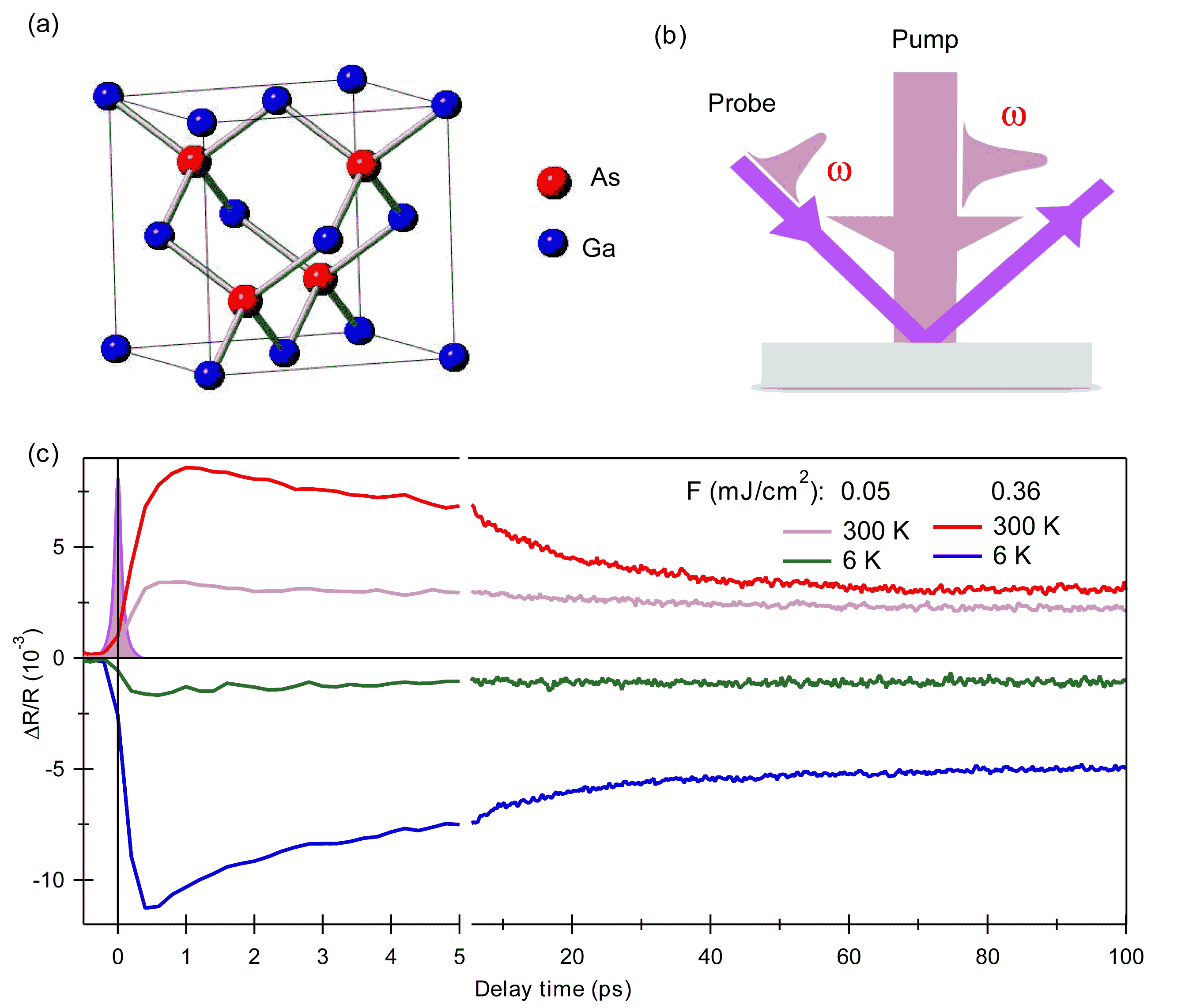}
   \caption{(a) Crystal structure of GaAs. (b) Mechanism of ultrafast time-resolved optical reflectivity. (c) Time-resolved optical spectra as a function of pump-probe delay at excitation densities of 0.05 and 0.36 $\rm{mJ/cm^2}$, at 300 K and 6 K, respectively. The zero time delay is determined by the the pink shaded curve, which is the second harmonic generation (FWHM of $\sim$120 fs) from the wave mixing $\omega_{pump}$ + $\omega_{probe}$.}
\end{figure}


 \begin{figure*}                                                                                                                                                  
   \includegraphics[width=2\columnwidth,scale=2,bb=0 0 3743 1877]{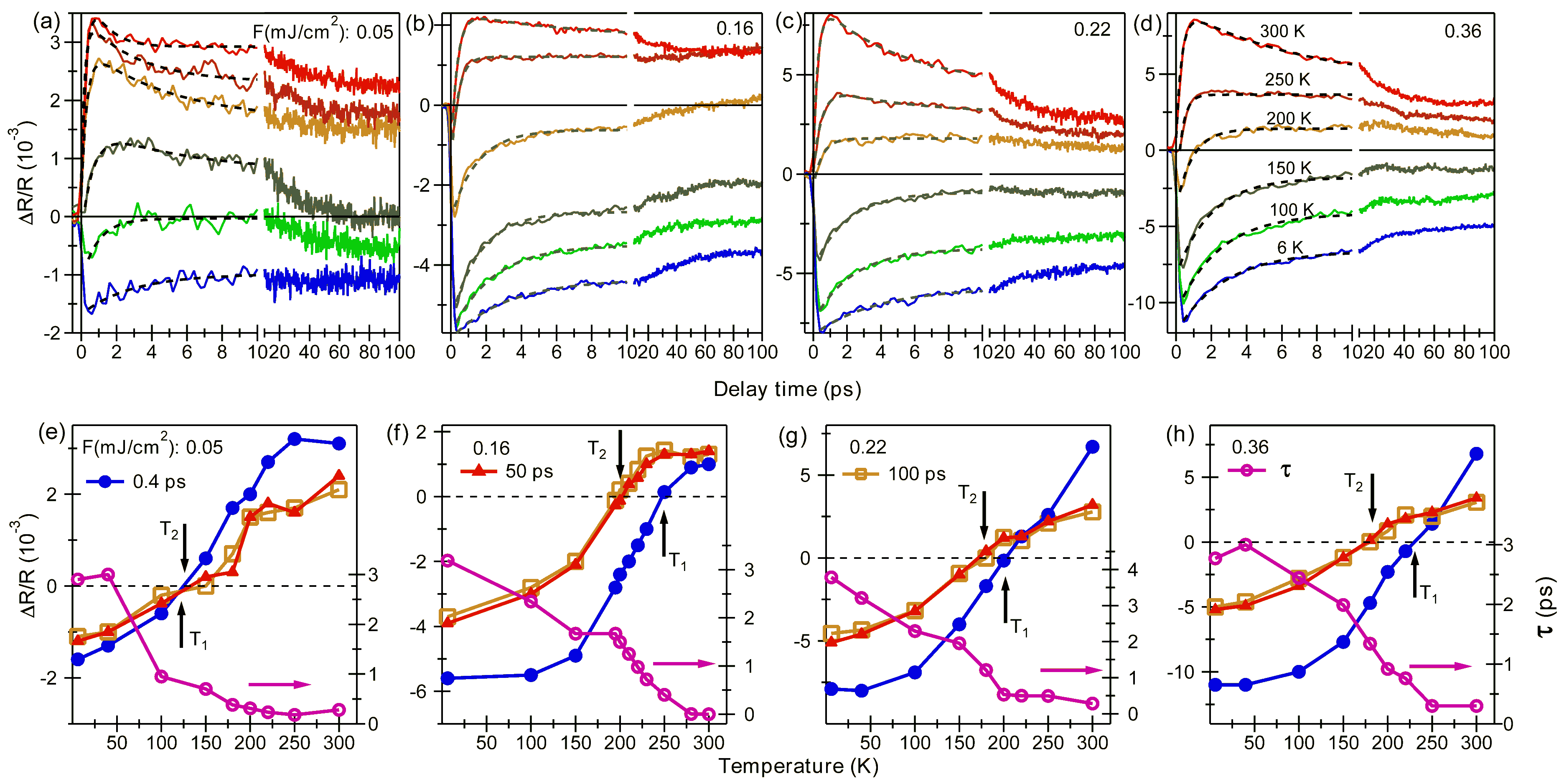}
   \caption{Time-dependent optical spectra of GaAs. (a-d) Reflectivity changes as a function
   of pump-probe delay measured at various temperatures,  for pump fluences at 0.05, 0.16, 0.22 and 0.36 $\rm{mJ/cm^2}$, respectively. The black dashed lines are exponential fits to 10 ps. (e-h) The variations of spectral intensity at the delay times of 0.4 ps (blue circles), 50 ps (red triangles) and 100 ps (brown open squares) as a function of temperature. The pink circles are decay time constant $\rm\tau$ for the 0.4 ps dip feature obtained by fitting curves in (a-d). The black arrows mark two critical temperatures T$_{1}$ and T$_{2}$, at which time-resolved spectra reverse from positive to negative.}
 \end{figure*}

\section{Results}
Figure 1(b) shows the mechanism of ultrafast time-resolved optical reflectivity. The system is excited by a pump beam to non-equilibrium states and tracked by a subsequent probe beam. The reflectivity change of probe is collected to study the transient dynamics of the system, such as light-induced phase transition, electron scattering process, collective modes and so on. Figure 1(c) shows time-resolved reflectivity changes induced by photoexcitation for different pump fluences at lattice temperatures of 6 K and 300 K, respectively. At room temperature, all transient spectra indicate a metallic feature in terms of $\Delta R/R>0$, owing to the above-gap excitation in which the pump photon energy (1.55 eV) is greater than the direct band-gap ($E_g\sim$ 1.42 eV) in GaAs. The absorption of pump photons generates a large number of hot free carriers, including
 free electrons in the conduction band and holes in the valence band in non-equilibrium states\upcite{Othonos-JAP-1998}, giving rise
 to a rapid increase of transient reflectivity\upcite{{Auston-SSE-1978}, {Shank-SSC-1978}}.
 As shown by the curves at 300 K in Fig. 1(c), a rise of the reflectivity is observed immediately after photoexcitation, followed by a peak feature around 1 ps, indicating a non-thermal carrier re-distribution. Subsequent transient reflectivity rapidly decreases, corresponding to a thermalize process, in which the electrons and holes thermalize among themselves into a Fermi-Dirac distribution, with temperatures that are typically higher than that of the lattice, in a few hundreds of femtoseconds primarily via carrier-carrier scattering\upcite{{Lin-APL-1987},{Nuss-PRL-1987}}. Within a couple of picoseconds the electrons and holes then thermalize with each other to reach a common temperature via electron-hole scattering\upcite{{Shah-1999}}. Then the transient reflectivity delays back to a positive constant from 40 ps to hundreds of ps. This corresponding to the cooling processes, in which the thermalized carrier distribution continues to cool to the lattice temperature by further interactions of the carriers with the lattice primarily via carrier-phonon scattering\upcite{{Cho-PRL-1990},{Kash-PRL-1985},{Kuznetsov-PRB-1995}}. The constant shows no change in our measurement for up to 300 ps, suggesting a stable transient metal stable phase. It is most likely due to a carrier heat diffusion process or a interband relaxation process, in which electron-hole pairs  recombine by either radiatively (emitting photons) on a time scale of hundreds of picoseconds to nanoseconds in direct gap semiconductors or non-radiatively, such as Auger recombination or carrier trapping, within tens to hundreds of picoseconds depending on the carrier density.

 More interestingly, on the other hand, we observe a sign reverse to negative for the transient spectra at 6 K ($\Delta R/R<0$) after photoexcitation, consisting of a well-defined dip feature at 0.4 ps and a negative constant from 40 ps to hundreds of ps, indicating a new non-equilibrium phase distinct from the transient metallic state at 300 K. The origin of this new transient phase will be discussed in the following section.

 In order to investigate the sign reverse process and corresponding dynamics in detail, we performed temperature-dependent experiments for pump fluence at 0.05, 0.16, 0.22 and 0.36 $\rm{mJ/cm^2}$, as depicted in Fig. 2(a-d). The peak feature at $\sim$ 1 ps and the positive constant around 300 K become suppressed with decreasing temperature and vanishes below $\sim$ 150 K. In the meanwhile, a negative dip at $\sim$ 0.4 ps starts emerging at $\sim$200 K (Fig. 2(b) and (c)) and gets enhanced in the cooling process, with a negative constant at lower temperature. As conveyed by Fig. 2(e-h), the changes of spectral intensity as a function of temperature are selectively extracted for delay times at 0.4, 50 and 100 ps. The curves for 50 and 100 ps are coincident with each other, confirming the constant feature and a long-surviving transient state. Two critical temperatures are identified by crossing to the zero baseline: Firstly, the negative dip at 0.4 ps $\Delta R/R<0$ (t = 0.4 ps) emerges, at T$_{1}$; Secondly, the damped negative constant $\Delta R/R<0$ (t $\geq$ 50 ps) forms below T$_{2}$. It is interesting to note that, in the region between T$_{1}$ and T$_{2}$, we find the change of reflectivity first produces a negative dip at 0.4 ps, which damps into a positive constant, suggesting a possibly composite state. To investigate the dynamics of the dip at 0.4 ps right after pump stimulation, we primely fit the relaxation processes to 10 ps by exponential functions, as indicated by the black dotted lines in Fig. 2(a-d). The decay time curves as a function of temperature $\tau$(T) in Fig. 2(e-h) show abrupt changes consistent with transitions at T$_{1}$ and T$_{2}$.
\begin{figure*}                                                                                                                                                          
\includegraphics[width=1.8\columnwidth,scale=1.8,bb=0 0 3379 2288]{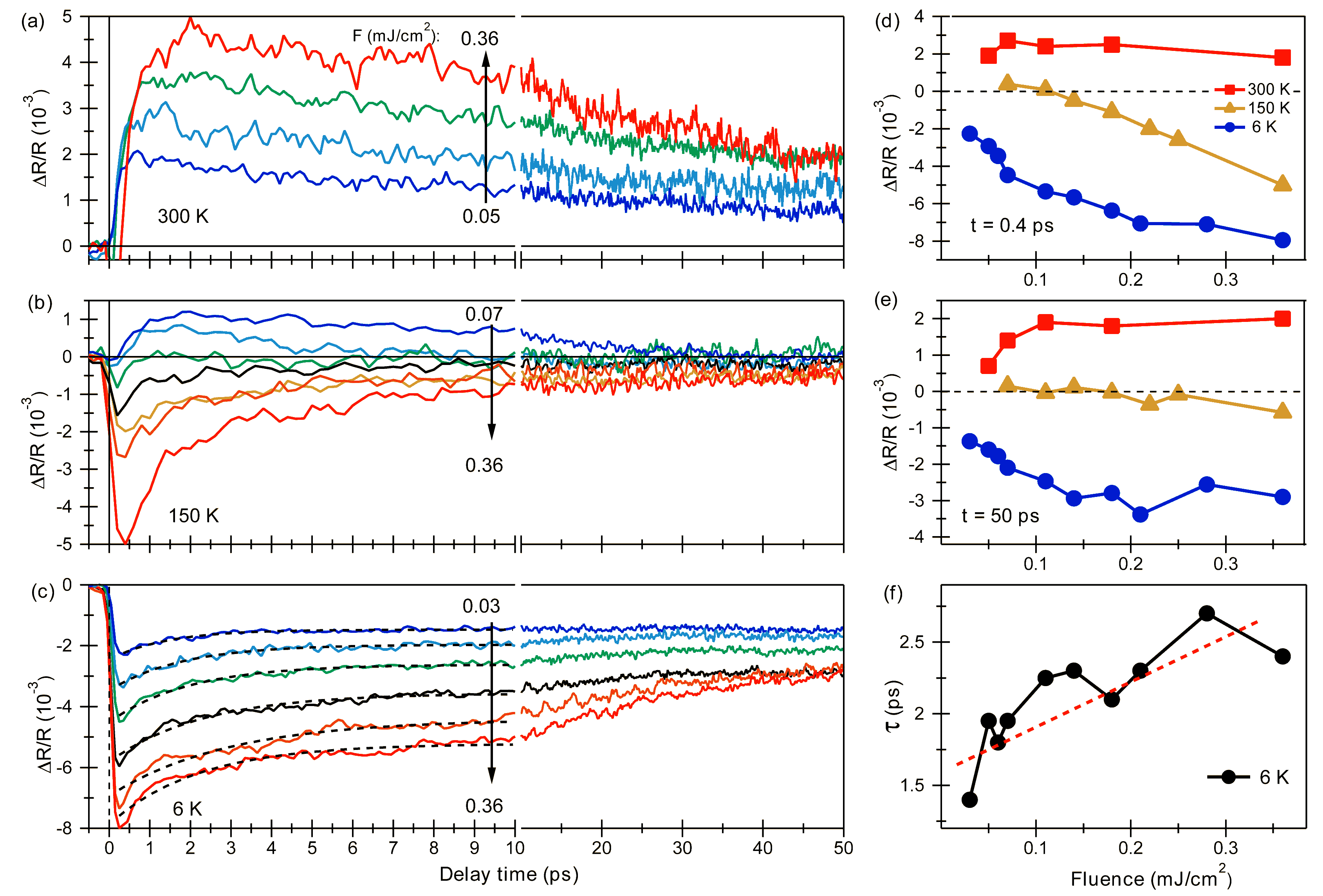}
\caption{Reflectivity spectra of GaAs after a series of excitation fluence at temperatures 300 K (a), 150 K (b) and 6 K (c). (d) and (e) Typical
 reflectivity changes at delay times 0.4 ps and 50 ps as a function of pump fluence, for different temperatures of 6 K (red squares), 150 K (brown triangles) and 300 K (blue circles), respectively. (f) Decay time constant of the dip feature at 0.4 ps at 6 K by exponential fits. The red line is used to guide the linear increment.}
\end{figure*}

 Figure 3 presents the influence of excitation density, by varying pump fluence from few percent to 0.36 mJ/cm$^2$, at fixed temperatures of 300, 150 and 6 K. For t$>$0, $\Delta R/R$ is constantly positive at 300 K (Fig. 3(a)) and increases with increasing pump fluence, while $\Delta R/R$ is all negative at 6 K (Fig. 3(c)) and shows a opposite behavior with increasing pump fluence. Remarkably, at 150 K (Fig. 3(b)), the increment of pump fluence induces a sign reverse of $\Delta R/R$ from positive to negative, suggesting a light-control of phase transition. Qualitatively analysis of the spectral intensities at 0.4 and 50 ps are shown in Fig. 3(d) and (e). Obviously, the change of reflectivity for 300 K gets enhanced below 0.11 mJ/cm$^2$ with a linear dispersion and saturates at higher pump fluence. At 150 K, the change of reflectivity starts a positive-negative reverse at 0.11 mJ/cm$^2$ in terms of the dip feature at 0.4 ps, and completely turns to negative above 0.22 mJ/cm$^2$ characterized by the constant feature around 50 ps. For 6 K, the first negative dip at 0.4 ps linearly increases its magnitude while the damped constant (around 50 ps) saturates above 0.15 mJ/cm$^2$. The decay time of the fast dynamics within 10 ps at 6 K linearly increases from 1.3 to 2.7 ps with increasing pump fluence, as indicated in Fig. 3(f).

\section{Discussion}

We now turn to the interpretation of the negative spectra ($\Delta R/R < $0) at cryogenic temperatures. One possibility is high-energy carriers by multi-photon absorption, similar to second harmonic pumping in Ref. \cite{Auston-SSE-1978}, which produces occupied high-energy conduction band states above the probe frequency, resulting in reduced imaginary part of the dielectric function and reduced reflectivity at the probe frequency\cite{Auston-SSE-1978}. However such a reason can be excluded since the lifetime of occupation of high-energy conduction band states is less than 2 ps as observed in Ref. \cite{Auston-SSE-1978}, due to fast intraband scattering, much less than the long-time constant feature for hundreds of ps observed in our experiments. The fact that the sign reverse is temperature-dependent at all excitation densities (Fig. 2(a)), also suggests an intrinsic phase transition rather than the multi-photon absorption mechanism.

Around room temperature, the direct band-gap in GaAs is $\sim$ 1.42 eV, above-gap excitation by 800 nm (1.55 eV) effectively produces hot free carriers, as shown in Fig. 4(a). The band-gap increases up to $\sim$ 1.52 eV upon cooling to 6 K, in which excitons play a major role after photoexcitation. As shown in Fig. 4(b) and (c), an exciton is a photon-excited heavy electron-hole pair glued by the Coulomb interaction, with the binding energy of 4.2 meV in GaAs\upcite{{Harrison-FCC-2016},{Frenkel-exciton},{Wannier-exciton}}. Excitons may be easily thermalized to free carriers at room temperature by absorbing lattice energy, which is much higher than the exciton binding energy. Without the Coulomb interaction and exciton contribution, the light absorption coefficient of semiconductors with a direct band-gap can be simply described by the function $\alpha(\hbar\omega) \propto \sqrt{\hbar\omega-E_{gap}}$, for $\hbar\omega > E_{gap}$. However, at cryogenic temperatures when the band-gap enlarges and the lattice energy becomes comparable or even smaller than the exciton binding energy, exciton stabilizes and contributes to a robust peak in the vicinity of the band edge\upcite{MDS-Ex-1962}, as shown by the solid line in Fig. 4(d). In contrast to free carriers, an exciton can be treated as an hydrogen-like bosonic particle\upcite{LV-SP-1968}, with exciton Bohr radius of 13 nm for GaAs. It is possible that excitons reduce the conductivity and optical reflectivity of GaAs to negative ($\Delta R/R<0$), which is typically an transient insulating behavior\upcite{Fehrenbach-PRL-1982}. Accordingly, the photon-induced transient state characterized by $\Delta R/R<0$ (t $>$ 0) can be attributed to an exciton insulating state, which has been observed before below 49 K\upcite{Amo-PRB-2006}.

Figure 4(e) summarizes the phase diagram of GaAs by extracting transition temperatures T$_1$ and T$_2$ from our results in Fig. 2. The dome-like phase
diagram consists of three regimes: (1) Above T$_1$, photoexcitation effectively produces free carriers, giving rise to positive $\Delta R/R$ and a free carrier metallic state; (2) Below T$_2$ is a transient exciton insulating state, in which the system holds a transient insulating property with negative $\Delta R/R$, due to the population and stabilization of excitons; (3) Between T$_1$ and T$_2$, pump pulses immediately stimulates a fast negative dip at $\sim$ 0.4 ps in $\Delta R/R$, which decays into a positive constant, noted as an exciton and free carrier composite state. The fact that the insulating state emerges when the metallic state is suppressed below T$_2$, as shown in Fig. 3(b), provides a possible way of control MIT by photoexcitation in semiconductors, as it has been realized in strongly correlated materials\upcite{ATCDW-PRL-2009,JC-PRL-2011,HY-PRB-2013,DP-NM-2007,RIR-PRL-2008}.
\begin{figure*}                                                                                                                                                    
\includegraphics[width=1.8\columnwidth,scale=1.8,bb=0 0 3614 1912]{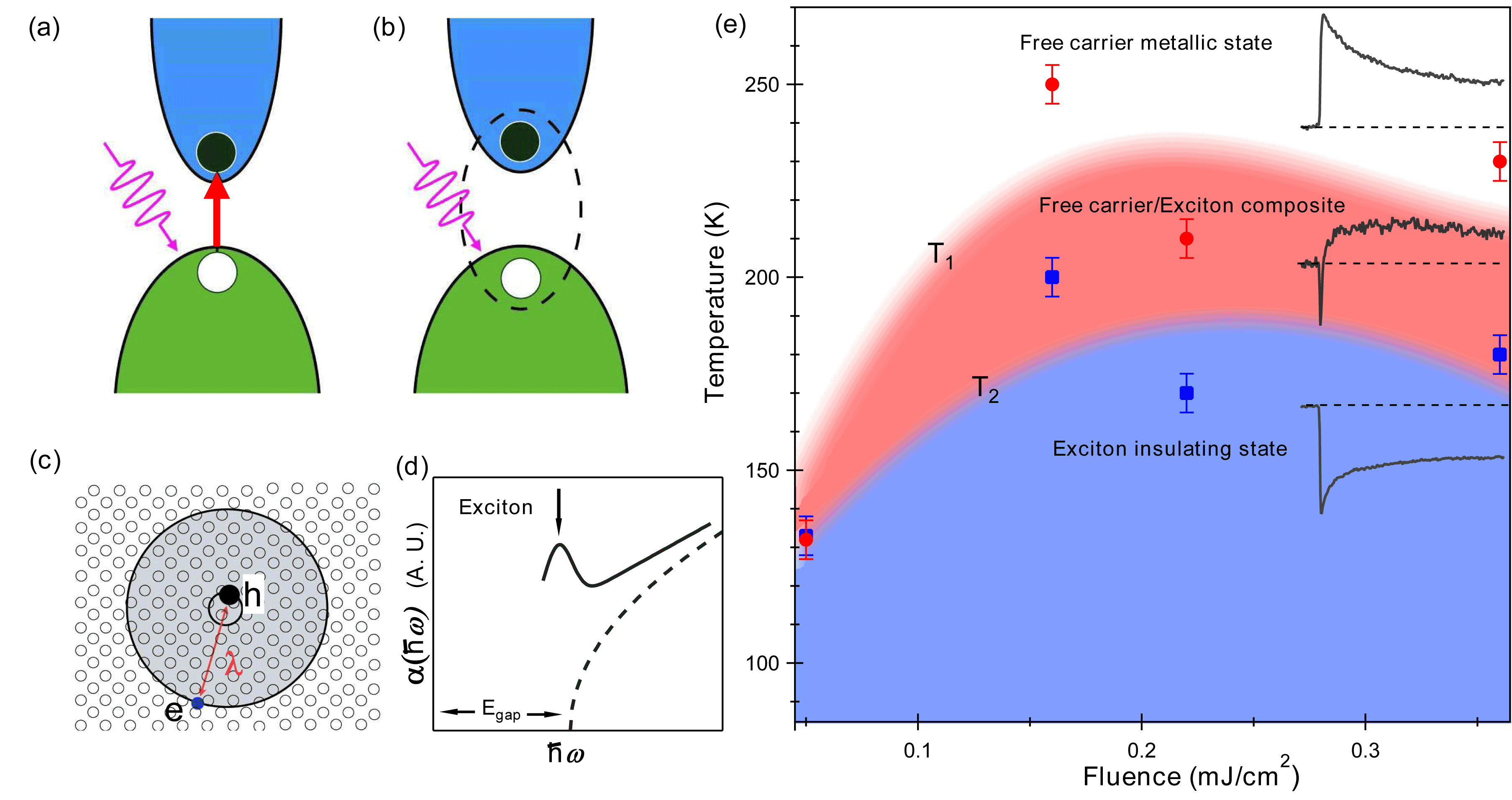}
\caption{Light-induced free carriers (a) and exciton (b) in a direct band-gap semiconductor. (c) A hydrogen-like free exciton in real space with the Coulomb interaction between hole and electron. $\lambda$ is the exciton Bohr radius (13 nm for GaAs). (d) Classic optical absorption coefficient $\alpha(\hbar\omega)$ without Coulomb interaction (dashed curve) and absorption for GaAs at cryogenic temperatures (solid curve), following Ref. \cite{MDS-Ex-1962}. The solid arrow marks the exciton absorption peak on the edge of the direct band-gap. (e) Transient phase diagram of GaAs. T$_1$ is the starting temperature of free carrier/exciton composite state, and T$_2$ is the exciton stable temperature. The inset curves are typical spectra in different phase regions.}
\end{figure*}


In our measurements, T$_1$ and T$_2$ are found to be as high as $\sim$ 230 K and $\sim$ 180 K, respectively, in sharp contrast to the exciton insulating temperature previously observed at 49 K\upcite{Amo-PRB-2006}. The difference is probably due to that highly intense pump pulse in our case induces high density of exciton, which elongates the exciton lifetime (Fig. 3(f)), and stabilizes the exciton insulating state, while low excitation density causes only uncorrelated excitons\upcite{RPP-Ex-2009}, yet theoretical work for such an light-enhanced exciton insulating state is required. The curves in composite state between T$_1$ and T$_2$, which consists of a fast negative dip and a flat positive constant, reminds us similar pump-probe reflectivity observed in high-$T_c$ cuprates, which is attributed to be the coexistence of pseudo-gap and superconducting gap\upcite{YHL-PRL-2008}.

In conclusion, we have investigated the transient dynamics in GaAs by ultrafast pump-probe optical reflectivity. Our results reveal that ultrafast photons effectively populate free carriers and excitons above T$_1$ and below T$_2$, giving rise to transient metallic and insulating states, respectively. We find that MIT can be induced by varying either temperature or excitation density. More quantum effects induced by the overlap of exciton wave functions at much higher exciton densities, such as biexciton, exciton Mott transition and Bose-Einstein condensation (BEC), require further ultrafast experiments and theoretical studies on semiconductors.

\section{Methods}
\subsection{Ultrafast time-resolved optical reflectivity}
Ultrafast time-resolved optical reflectivity based on femtosecond laser is employed to elucidate the electronic dynamics of n-type GaAs(100).
The optical pulses were generated by a Ti:sapphire laser amplifier system (repetition rate of 1 KHz, pulse duration at 35 fs
and central wavelength of 800 nm) and split by a beam splitter into intense pump and weak probe pulses. Both pump and probe were focused onto the sample, with spot sizes at $\sim$ 0.4 mm (pump) and $\sim$ 0.2 mm (probe) in diameter. The reflected probe signal was collected by a Si-based detector and a lock-in amplifier. The temporal evolution of the reflectivity change ($\Delta R$) was measured by scanning the delay time between pump and probe pulses, using a motorized delay line. The sample was mounted on a cryostat with a temperature sensor embedded close by, allowing a precise control of temperature in the range of 6$-$300 K.

\section{Acknowledgements}
We are grateful to acknowledge support from the National Natural Science Foundation of China (Grant No. 51275012) and NSAF of China
(Grant No. U1530153). H. Y. Liu thanks the Sea Poly Project of Beijing Overseas Talents (SPPBOT).

\section{Author Contributions}
X.C., S.B and H.Y. contributed to the proposal for experiment. X.C., X.Z., G.P. and H.S. carried out the experiment under the supervision of S. B. and H.Y.. F.L. provided the sample. X.C. and H.Y. analyzed the data and wrote the manuscript, with the contributions from all other authors.

\section{Competing Interests}
The authors declare no competing interests.

\end{document}